\documentclass[osajnl,twocolumn,showpacs,superscriptaddress,10pt]{revtex4-1} 
\usepackage{amsmath,amssymb,graphicx}
\begin{document}

\title{Optical levitation of microdroplet \\containing a single quantum dot}

\author{Yosuke Minowa}
\email{Corresponding author:minowa@mp.es.osaka-u.ac.jp}
\author{Ryoichi Kawai}
\author{Masaaki Ashida}

\affiliation{Graduate School of Engineering Science, Osaka University, Toyonaka, Osaka 560-8531, Japan}

\begin{abstract}We demonstrate the optical levitation or trapping in helium gas of a single quantum dot (QD) within a liquid droplet. Bright single photon emission from the levitated QD in the droplet was observed for more than 200 s. The observed photon count rates are consistent with the value theoretically estimated from the two-photon-action cross section. This paper presents the realization of an optically levitated solid-state quantum emitter. 
\end{abstract}


\maketitle 

Optical traps, optical tweezers, or optical dipole traps are ubiquitous techniques\cite{grier_revolution_2003,ashkin_acceleration_1970} that are widely used in fields such as molecular biology, chemistry, and physics. These techniques have formed the foundation of single-molecule research. For example, micrometer-sized beads optically trapped in room temperature liquid have been used as a handle to manipulate DNA, RNA, and molecular motors\cite{grier_revolution_2003}. Optical trapping is also indispensable for capturing, aligning and cooling cold atoms\cite{mckeever_state-insensitive_2003}. Far-off-resonance optical dipole traps, together with optical cavities, provides a rich playground for studying the interaction between  single atoms and single photons\cite{birnbaum_photon_2005}. The longest trapping lifetime reported to date for a single atom is  $\sim100$ s\cite{he_extending_2011}, which is sufficiently long to demonstrate coherent state manipulation in the strong coupling regime\cite{mckeever_state-insensitive_2003}. 

The attainment of the strong-coupling regime requires an optical cavity with a high quality-factor and a small mode-volume, a large transition dipole moment and precise free-space positioning of the dipole with respect to the cavity. The requisite free-space positioning has been attained using an optical dipole trap\cite{nusmann_submicron_2005}. Thus, combining an optical dipole trap with a much larger transition dipole moment should result in  ultra strong coupling, or even deep strong coupling, enabling us to further investigate novel quantum-optics phenomena\cite{stassi_spontaneous_2013, de_liberato_light-matter_2014}. At present, the ultrastrong coupling regime is achieved in superconducting quantum circuit systems and condensed-matter systems\cite{niemczyk_circuit_2010,scalari_ultrastrong_2012}. 

A large transition dipole moment is available from collective excitations in condensed matter systems, and  solid-state semiconductor nanocrystal (i.e., QDs), which are very stable quantum emitters, exhibit such collective excitation in the form of excitons\cite{trindade_nanocrystalline_2001}. Through the quantum confinement effect, the material from which QDs are made and their shape and size determine the QD energy levels. The strong confinement provided by QDs  also enhances the transition dipole moment\cite{kayanuma_quantum-size_1988}. Of the many types of the QDs, chemically synthesized colloidal QDs are well-suited for this application as their size can be controlled with precision and they can be mass produced\cite{trindade_nanocrystalline_2001}. Colloidal QDs act as a single isolated entity and several groups have reported on optically manipulating and trapping colloidal QDs in liquid\cite{jauffred_three-dimensional_2008}. However, no report that demonstrates optical levitation\cite{neukirch_observation_2013, arita_laser-induced_2013} or optical trapping of a single QD in gas or vacuum has yet appeared.

To obtain stable optical levitation of a single QD, a major obstacle is the small nonresonant polarizability of the QD. In the Rayleigh regime $r \ll \lambda$, the nonresonant polarizability $\alpha$ of a particle with radius $r$ is proportional to the volume\cite{jauffred_three-dimensional_2008} $\alpha \propto r^3$. Consider CdSe/ZnS core/shell colloidal QDs as an example; the typical polarizability in the transparent region is $\alpha/\epsilon_0\sim 2.6 \times 10^{-25}\, \mathrm{m}^{3}$. If a high-power (1 W), 785 nm Gaussian beam is focused via an "ideal" lens\cite{tey_interfacing_2009}, the trap depth would be very close to room-temperature thermal energy, which is insufficient for stable trapping. This shallow trap depth shows strong contrast with the recent successful stable levitation experiments of submicro particles\cite{gieseler_subkelvin_2012, kiesel_cavity_2013-1}, whose sizes are one or two orders of magnitude larger than the size of the QD. The most plausible way to circumvent this problem is to embed the single QD into a larger volume of transparent material. In the present study, we demonstrate that a strong single-beam gradient force can optically levitate a single CdSe/ZnS core/shell QD embedded in a micrometer-sized liquid droplet in helium gas. The microdroplet is easy to produce and handle and is highly symmetric, making it well suited for stable optical levitation. A charged microdroplet containing many QDs has previously been electrodynamically trapped\cite{schafer_quantum_2008}. The optical levitation technique demonstrated here is not restricted to the trapping of charged particles. The technique is also applicable to the trapping of neutral particles.

\begin{figure}
\begin{center}
\includegraphics[scale=0.6]{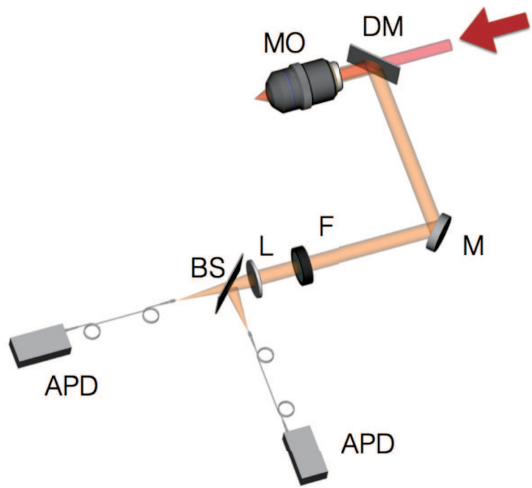}
\end{center}
\caption{Schematic of the experimental. Near-infrared light from a continuos-wave Ti:sapphire laser was focused using a microscope objective (MO). A dichroic mirror (DM) reflected two-photon-excited fluorescence from a levitated QD. To suppress stray light, we spectrally filtered the fluorescence using bandpass filters (F). Next, the fluorescence was focused by a lens (L) and split into two beams of equal intensity by a 50/50 beam splitter (BS). Each beam was coupled into multimode optical fibers and detected by avalanche photodiodes (APDs) or a spectrograph.
\label{fig:setup}}
\end{figure}

\begin{figure}
\begin{center}
\includegraphics[scale=0.2]{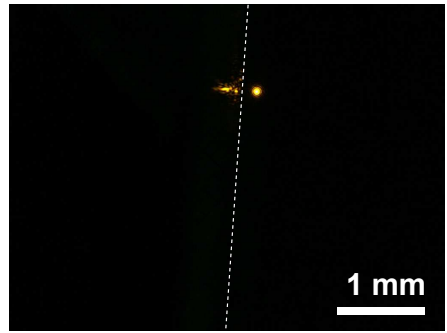}
\end{center}
\caption{Levitated microdroplet. Strong scattering of the trapping laser from the levitated microdroplet. Dotted line indicates the inner surface of one of the coverslips forming the sample chamber. The objective lens is on the left in the image. The O-ring and the other coverslip are removed for image clarity. After tentatively blocking the trapping laser, the strong scattering disappeared, confirming the escape of the microdroplet from the optical trap (see Media 1).
\label{fig:levitation}}
\end{figure}

Using a strongly focused near-infrared laser beam with a 785-nm wavelength, we optically levitated an ethanol microdroplet containing a chemically synthesized CdSe/ZnS QD with an emission wavelength of 640 nm and a 3.15-nm radius (Sigma Aldrich) at room temperature. The QDs were diluted to $10^{-3}$ in ethanol. This solution was sonicated in an ultrasonic bath for 30 min before use. Figure \ref{fig:setup} shows the experimental setup. The laser power after the objective lens (Olympus, UPLFLN60X) was 480 mW, which is sufficient for the stable optical levitation of the micrometer-sized ethanol droplet. For example, the trap depth for the microdroplet with the diameter of 1 $\mu$m exceeds $10^3$ times of the room-temperature thermal energy in our experimental condition, while the stable position of the microdroplet is calculated to be slightly shifted due to the scattering force\cite{nieminen_optical_2007}. The ethanol microdroplets were introduced into the sample chamber via an ultrasonic nebulizer\cite{neukirch_observation_2013}. The sample chamber was formed by two 170-$\mu$m-thickness coverslips separated by a rubber O-ring (mounted with vacuum grease to prevent airflow). Inside a sample chamber filled with helium gas, the trapping lifetime of the microdroplet exceeded several dozen minutes. The use of the helium gas is desirable to suppress the photo-oxidation of the QDs. We confirmed the optical levitation via the strong scattering of the laser from the microdroplet as shown in Fig. \ref{fig:levitation}. If we tentatively blocked the laser, the strong scattering ceased thereafter, confirming the escape of the microdroplet from the optical trap (see Media 1). The optical absorption of the ethanol droplet has a negligible effect on the quantum efficiency of the fluorescence of the single QD (the absorption coefficient of liquid ethanol is $\sim 10^{-2} \,\mathrm{cm}^{-1}$ at 785 nm and $\sim 10^{-3} \,\mathrm{cm}^{-1}$ across the entire visible spectrum). 

\begin{figure}
\begin{center}
\includegraphics[scale=0.7]{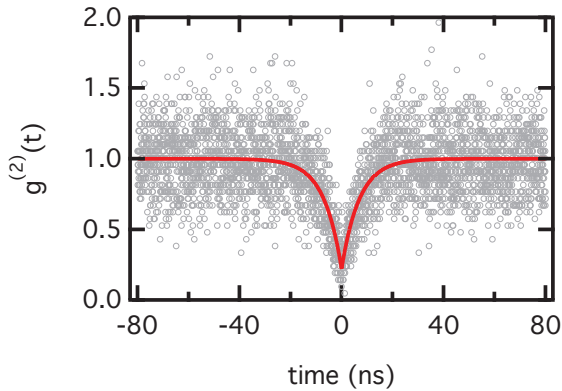}
\end{center}
\caption{Single levitated solid-state quantum emitter. Normalized second-order correlation function $g^{(2)}(t)$ of two-photon-excited single-QD fluorescence.  The solid curve is a fit to a single exponential and gives a time constant of 6.9$\pm$ 0.3 ns. The result $g^{(2)}(t=0)=0.22$ confirms that the levitated-QD is a single-photon source.
\label{fig:antibunching}}
\end{figure}

Through two-photon absorption, the trapping laser excited the optically levitated QD inside the droplet. CdSe/ZnS QDs have a large two-photon-action cross section, which is the product of the nonlinear two-photon-absorption cross section and the fluorescence quantum efficiency\cite{larson_water-soluble_2003}. Based on the reported two-photon-action cross section 2,000 - 50,000 Goeppert-Mayer units (1 GM = $10^{-50}$ cm$^4$ s/photon), we estimated the maximum rate of photon emission to be $10^7 \sim 10^8$ photons/s. The two-photon-excited fluorescence was split by a 50/50 beam splitter into two beams and coupled into multimode optical fibers (see Figure \ref{fig:setup}). The small fiber core (diameter 105 $\mu \mathrm{m}$) provides a confocal pinhole and eliminates a large fraction of the stray light. The output of the two fibers was directed onto single-photon-counting avalanche photodiode modules (Excelitas Technologies, SPCM-AQRH-14). The Hanbury Brown and Twiss setup allows us to investigate the photon statistics of the levitated photon source. Figure \ref{fig:antibunching} shows the measured two-photon coincidence counts $g^{(2)}(t)$ fit to an exponential function (solid line). If the levitated QD emits a single photon at a time, the photon coincidence at $t =0$ would ideally go to zero. The observed small value of $g^{(2)}(t)=0.22$ indicates that the optically levitated QD behaved as a single-photon source. The small offset remaining at $t =0$ may be due to stray light and the biexciton cascade decay\cite{nair_biexciton_2011}. The time constant $\tau=6.9$ ns obtained from the fit is related to the fluorescence lifetime $\tau_\mathrm{decay}$ and the excitation rate $\Gamma$ by $\dfrac{1}{\tau}=\dfrac{1}{\tau_\mathrm{decay}}+\Gamma$ and is consistent with the previous studies\cite{michler_quantum_2000}. Almost half of the trapped droplets showing photoluminescence contain only a single QD, whereas the rest contain more than a single QD. We can trap a several droplet showing photoluminescence in a few minutes.

\begin{figure}
\begin{center}
\includegraphics[scale=0.4]{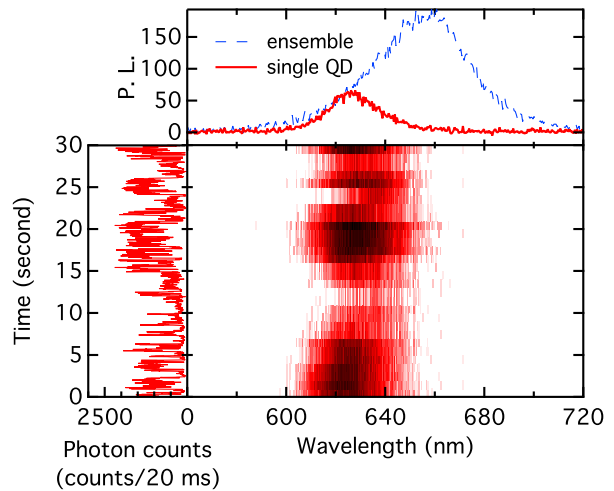}
\end{center}
\caption{Spectral and intensity fluctuations in fluorescence from levitated QD. (lower right panel) Time-dependent fluorescence spectra from a single levitated QD. Integration was 1 s. (upper right panel) The red curve is a typical fluorescence spectrum extracted from the lower panel at $t=5$ s. For reference, the dotted blue curve shows the fluorescence spectrum from the QD ensemble. (left panel) Time-dependent fluorescence intensity detected by an avalanche photodiode with 20 ms bin time.
\label{fig:timetrace}}
\end{figure}

The lower right panel of Fig. \ref{fig:timetrace} shows the temporal dynamics of two-photon-excited fluorescence from a levitated QD. After recording the two-photon coincidence counts and ensuring the single-photon emission, one of the avalanche photo diodes was replaced by a spectrograph and the fluorescence was recorded with a 1 s integration time. The fluorescence line width is similar to the previously reported value\cite{shi_excitation_2014} and much narrower than that of the QD ensemble (see dotted blue curve in upper right panel of Fig. \ref{fig:timetrace}), which indicates that inhomogeneous broadening has been eliminated. However, this line width $15.2 \pm 0.2$ nm may still include the effect of rapid spectral diffusion within the integration time\cite{coolen_emission_2009}. The fluorescence spectrum of the QD ensemble was derived from a trapped droplet containing many QDs. Moreover, the photoluminescence intensity evidently fluctuated during acquisition. To further examine the temporal change of the fluorescence dynamics, we recorded it using the avalanche photodiode (left panel of Fig. \ref{fig:timetrace}). The integration time was 20 ms/bin, and data were recorded over 200 s. The fluorescence intensity from the levitated QD strongly fluctuated on a sub-second-scale, reflecting the Brownian motion of the QD within the liquid droplet and the blinking characteristic, which is expected from single-QD fluorescence\cite{michler_quantum_2000,hohng_near-complete_2004}. The effective excitation power weakly depends on the location of QD inside the microdroplet, as our confocal detection volume size is similar to the size of the microdroplet. Therefore, the Brownian motion causes the time-dependent effective excitation power change, leading to the continuous intensity fluctuation. The measured maximum photon counting rate was about $2\times 10^5$ counts/s. Considering our detection efficiency of at most 6 \%, we calculate that the photon emission rate during optical levitation exceeded $10^6 \sim 10^7$ photons/s. Such a high photon-emission rate is consistent with the value estimated above from the two-photon-action cross section and provides convincing evidence of bright single-photon emission. Typically, we observed bright single-photon emission for 100-200 s. Even after the stop of the bright single-photon emission, the droplet was still trapped. Although one may expect an enhanced spontaneous emission rate through the Purcell effect\cite{purcell_spontaneous_1946} (due to the presence of the ethanol micro-droplet), we saw no significant spontaneous emission enhancement. The fluorescence spectra also clearly show that no cavity mode is present.

Overall, our results shows that by exploiting the optical gradient force on a microdroplet containing a single QD, a focused laser beam can position a QD at a given position in free space. To improve on these results, the internal temperature and center-of-mass motion of the levitated QD could be reduced by buffer-gas cooling\cite{doret_buffer-gas_2009}. In addition, stable optical levitation of a bare QD would become straightforward if the center-of-mass motion is reduced by cooling to sub-Kelvin regime. The current approach is also readily extendable to multistep optical levitation, whereby the QD-microdroplet system is optically levitated, followed by cooling to reduce motion and subsequent levitation of a bare QD in a shallow optical trap. Based on our current experimental parameters, we estimate that the optical trap depth for a bare QD would be the half of the room temperature thermal energy. Therefore, the proposed multistep optical levitation is possible with the reported feedback cooling technique of micro and nanoparticles down to the sub-Kelvin regime\cite{li_millikelvin_2011,gieseler_subkelvin_2012}.

Finally, the experimental methods developed in this study can lead to optical levitation of a number of fascinating nanomaterials and their concomitant isolation from the substrate or matrix. Because of the large surface-to-volume ratio, the properties of nanomaterials tend to depend strongly on their environment\cite{park_near-unity_2011, zhao_effect_2012}, and environment-induced effects often impede the detailed study of the nanomaterials. Thus, optical levitation in gas or vacuum would provide an ideal platform to investigate nanomaterials and should lead to an improved understanding of their intrinsic properties\cite{neukirch_observation_2013,arita_laser-induced_2013}.

The authors thank H. Ishihara, M. Kumakura, Y. Moriwaki, and T. Torimoto for their useful discussions.  This study was partially supported by KAKENHI from the MEXT of Japan.


%

\end{document}